%% file: main_v2_uncolored.tex
\documentclass[showkeys,floatfix,noshowpacs, preprintnumbers, aps,superscriptaddress,pra,10pt,twocolumn,longbibliography,nofootinbib]{revtex4-2}

\input{preamble}

\usepackage{xcolor}

\usepackage{titlesec}
\titlespacing*{\paragraph}{\parindent}{*1}{0.1em}[0pt]
\begin{document}

\title{Questioning the absoluteness of free choices when internalized in Wigner's friend scenarios}
\author{Laurens Walleghem}\thanks{\newline Email: \href{laurens.walleghem@york.ac.uk}{laurens.walleghem@york.ac.uk}}
\affiliation{Department of Mathematics, University of York, Heslington, York YO10 5DD, United Kingdom}
\affiliation{International Iberian Nanotechnology Laboratory (INL), Av. Mestre Jos\'{e} Veiga, 4715-330 Braga, Portugal}\date{October 2024}

\begin{abstract} 
Wigner's thought experiment illustrates quantum theory's measurement problem  by considering an observer who measures a quantum system inside a sealed lab, modeled unitarily by an outsider.
Recent extensions of this thought experiment, referred to as extended Wigner's friend arguments, question how different observers can reason consistently about each other in quantum setups, and challenge the absoluteness of the outcome value obtained by the friend under a notion of locality.
In this work, we present an argument against the absoluteness of free choices under the same notion of locality, using an extended Wigner's friend scenario based on the Pusey--Barrett--Rudolph theorem. Similar arguments based on other contextuality or nonlocality models are possible.
\end{abstract}

\date{\today}

\maketitle

\section{Introduction} \label{sec:intro}
One of the most prominent, longstanding challenges in quantum theory concerns the measurement problem~\cite{Maudlin1995,Pitowsky2007,BubPitowsky2010,brukner2017quantum,schlosshauer2005decoherence,legget20005measurement,Hance_2022,Briggs:2013qxp}.
Wigner's thought experiment~\cite{wigner1995remarks} illustrates this problem by considering an observer, referred to as Wigner's friend, who performs a measurement inside a sealed lab.
Wigner, who remains outside the sealed lab, is a \textit{superobserver} for the friend and describes the lab's evolution as a unitary interaction of quantum systems.
Therefore, Wigner claims that the friend is in a superposition of seeing different outcomes, at odds with the friend's claim of obtaining a single-valued outcome.
Even though Wigner and his friend disagree on whether, or when, the measurement collapse happens, there is no immediate contradiction concerning the actual outcome value of the friend's measurement.
Recent extensions of this thought experiment~\cite{brukner2017quantum,frauchiger2018quantum,bong2020strong,cavalcanti2021implications,haddara2022possibilistic,guerin2020no,utreras2023allowing,haddara2024local,szangolies2023quantum,walleghem_refined_2024,walleghem2023extended,walleghem2024connecting,walleghem2025extendedwignersfriendnogo,vilasini2019multi,vilasini2022general,ormrod2023theories,leegwater2022greenberger,schmid2023review,brukner2018no,ormrod2022no,nurgalieva2018inadequacy,montanhano2023contextuality,zukowski2021physics,walleghem2023extended,montanhano2023contextuality,guerin2020no,ying2023relating,schmid2023review,walleghem2025wignersfriendsblackhole}, referred to as extended Wigner's friend (EWF) arguments, provide strong no-go theorems on the nature of the actual outcome value obtained by the friend.  
One such celebrated result is the Local Friendliness (LF) no-go theorem~\cite{bong2020strong,cavalcanti2021implications,haddara2022possibilistic}, which derives a contradiction under the assumptions of the Absoluteness of Observed Events---that outcomes of performed measurements are absolute---and Local Agency---that free choices (interventions) can only be correlated with events in their future lightcone. 

In this work, we provide an extended Wigner's friend argument against the absoluteness of free choices, under the same Local Agency assumption.  
A free choice is usually taken as exogenous; by observing an observer who makes a free choice in a sealed lab, we will allow a superobserver to model this free choice in a particular internalized way, while it may still seem exogenous to the first observer. 
This argument therefore suggests that resolutions of extended Wigner's friend arguments that refute absoluteness should also incorporate a relational nature of free choices, or at least invites to rethink the (exogenous) concept of free choice and whether it can be internalized into a physical description.
The scenario we present is based on the Pusey--Barrett--Rudolph (PBR) scenario~\cite{pusey2012reality}, which tackled questions about the reality of the quantum state, and our extension can be seen as an entanglement swapping involving Bell nonlocality~\cite{bell1964einstein,Bell_1976,brunner_bell_2014}. Similar arguments can also be made using scenarios other than the PBR one.

The rest of this work is organized as follows. After reviewing the Local Friendliness no-go theorem and Pusey--Barrett--Rudolph theorem in \cref{sec:recap}, we present our main results in \cref{sec:LF_free_choices}. We end with a brief discussion in \cref{sec:discussion}.

\section{A short recap of Local Friendliness and Pusey--Barrett--Rudolph}  \label{sec:recap}

\subsection{The Local Friendliness no-go theorem}

The Local Friendliness scenario of Refs.~\cite{bong2020strong,haddara2022possibilistic} involves a Bell nonlocality set-up with two friends, i.e. two observers who are modeled unitarily by their respective superobservers.

Specifically, two friends, Alice and Bob, reside in sealed labs.
They share a bipartite system in the state $\rho_{RS}$, and each perform a measurement on their part of the system, modeled unitarily as $U_A,U_B$ by their respective superobservers Ursula and Wigner.
Next, Ursula and Wigner each make a choice $x,y$.
For $x=0$ Ursula asks Alice for her outcome $a$, whereas for $x \neq 0$ Ursula instead performs another operation on Alice's lab. 
For example, an operation that Ursula can perform, used to prove the LF no-go theorem, consists of undoing Alice's measurement by applying the inverse unitary $U_A^\dagger$ to her lab, resetting her lab back to its initial state and thus erasing the memory record of Alice's outcome, followed by a measurement on $R$.
Wigner acts analogously on Bob's lab.

This scenario can be used to prove
a no-go theorem against the Local Friendliness assumptions, which consist of Absoluteness of Observed Events and Local Agency~\cite{cavalcanti2021implications}.

\begin{assump}[Absoluteness of Observed Events (AOE)] \label{assumption:AOE}
An observed event is an absolute single event, not relative to anything or anyone. 
\end{assump}

\begin{assump}[Local Agency] \label{ass:LA}
    The only relevant events correlated with an intervention are in its future light cone.
\end{assump}

\begin{theorem}[LF no-go theorem~\cite{bong2020strong,haddara2022possibilistic,cavalcanti2021implications}] \label{th:LF}
If a superobserver can perform arbitrary quantum operations on an observer and its environment, then no physical theory can satisfy Local Friendliness.
\end{theorem} 

Specifications for the above scenario to prove the no-go theorem are such that the empirically obtained correlations\footnote{Following Refs.~\cite{bong2020strong,walleghem2024connecting,walleghem2025extendedwignersfriendnogo}, we denote empirical probabilities by $\wp$, to distinguish them from probabilities $p$ which are required to exist by principles such as AOE. To clarify, in this work we did not actually perform any experiments, and the empirical probabilities refer to probabilities concerning variables that remain accessible after the protocol in this thought experiment. Were one to implement the thought experiment in an actual experiment, they would be truly empirical probabilities.} $\wp(a,b|x{=}0,y{=}0)$, $\wp(a,w|x{=}0,y{=}1)$, $\wp(u,b|x{=}1,y{=}0)$, $\wp(u,w|x{=}1,y{=}1)$ cannot arise from a single global distribution $p(a,b,u,w)$.
These can be obtained by starting from an entangled bipartite state $\rho_{RS}$, Ursula and Wigner undoing the measurements of Alice and Bob followed by measurements on $S$ and $R$, with the measurements performed by the four agents violating Bell locality on the state $\rho_{RS}$.

\subsection{The Pusey--Barrett--Rudolph argument}

The Pusey--Barrett--Rudolph theorem~\cite{pusey2012reality} is a result against an epistemic reading of the quantum state within the \textit{ontological models framework}.
In this framework, a physical system is associated with one or more ontic states $\lambda$ about which a measurement can reveal certain information. 
Now it may occur that full knowledge of a preparation $P$ of the system does not specify a unique ontic state $\lambda$, but only a distribution $\mu_P(\lambda)$ over such ontic states, referred to as the \textit{ontic distribution}.
A measurement $M$ is represented by a probability distribution $p(\xi|\lambda,M)$ over its outcomes $\xi$ for each ontic state $\lambda$, and the empirical probability of obtaining $\xi$ upon a measurement $M$ given preparation $P$ is given by $p(\xi|M,P) = \sum_\lambda p(\xi|M,\lambda)\mu_P(\lambda)$, where the sum is replaced by an integral for continuous $\lambda$.

The question of whether the quantum state is real or not can be given the following interpretation in this framework.
The quantum state is said to be real, or \textit{ontic}, if each pair of distinct, pure quantum states $\ket{\psi_0},\ket{\psi_1}$ have non-overlapping distributions, $\mu_{\psi_0}(\lambda) \mu_{\psi_1}(\lambda) = 0$.
The quantum state is said to be \textit{epistemic} if it is not ontic, i.e. if there exists a pair of operationally distinguishable quantum states  $\ket{\psi_0},\ket{\psi_1}$ that have overlapping ontic distributions, $\mu_{\psi_0}(\lambda) \mu_{\psi_1}(\lambda) > 0$.
A quantum state being epistemic captures the idea that it represents mere knowledge.

The Pusey--Barrett--Rudolph argument against epistemic states in this framework goes as follows, following Ref.~\cite{pusey2012reality} closely.
Consider two copies of a preparation device that operate independently and can each prepare a quantum system in the state $\ket{0}$ or $\ket{+}$.  
Here $\ket{0},\ket{1}$ is a basis of a qubit, $\ket{+}=(\ket{0}+\ket{1})/\sqrt{2}$ and for later convenience we also define $\ket{-}=(\ket{0}-\ket{1})/\sqrt{2}$.
The possible preparations of the two devices jointly are thus $|0\rangle \otimes|0\rangle,|0\rangle \otimes|+\rangle,|+\rangle \otimes|0\rangle \text { and }|+\rangle \otimes|+\rangle$.

Beyond the ontological models framework assumption $(i)$, it is assumed that $(ii)$ a form of \textit{preparation independence} holds~\cite{leifer2014quantum,Mansfield_2016}, i.e., that the joint ontic distribution of multipartite preparations leading to product quantum states factorizes, $\mu_{P_A P_B}(\lambda_A,\lambda_B) = \mu_{P_A}(\lambda_A)\mu_{P_B}(\lambda_B)$. Note that this property is only required for product states, and need not be assumed for entangled states.  

Next, an entangled measurement is performed upon the two systems, which projects onto the four orthogonal states 
\begin{equation} \label{eq:definition_xi_PBR}
\begin{aligned}
& \left|\xi_1\right\rangle=\frac{1}{\sqrt{2}}(|0\rangle \otimes|1\rangle+|1\rangle \otimes|0\rangle), \\
& \left|\xi_2\right\rangle=\frac{1}{\sqrt{2}}(|0\rangle \otimes|-\rangle+|1\rangle \otimes|+\rangle), \\
& \left|\xi_3\right\rangle=\frac{1}{\sqrt{2}}(|+\rangle \otimes|1\rangle+|-\rangle \otimes|0\rangle), \\
& \left|\xi_4\right\rangle=\frac{1}{\sqrt{2}}(|+\rangle \otimes|-\rangle+|-\rangle \otimes|+\rangle).
\end{aligned}
\end{equation}
If the ontic distributions of the quantum states $\ket{0},\ket{+}$ overlap, then there is a nonzero probability $q^2 > 0$ that an ontic state $\lambda$ is prepared that is compatible with all possible preparations listed above.
Now each of these states $\ket{\xi_i}$ is orthogonal to a different one of the given preparations, implying that the corresponding measurement outcome $\xi_i$ has probability zero of occurring when given the prepared state orthogonal to $\ket{\xi_i}$.
Namely, $\ket{\xi_1}$ is orthogonal to $\ket{0}\otimes \ket{0}$,  $\ket{\xi_2}$ is orthogonal to $\ket{0}\otimes \ket{+}$,  $\ket{\xi_3}$ to $\ket{+}\otimes \ket{0}$ and  $\ket{\xi_4}$ to $\ket{+}\otimes \ket{+}$.
This fact immediately yields the contradiction; when given such $\lambda$ compatible with all possible preparations, the measurement device is unsure which preparation had been performed, and thus runs the risk of giving a measurement outcome for which the actual preparation had probability zero of occurring. 

This shows that the quantum states $\ket{0},\ket{+}$ must have zero overlap in their ontic distributions. A generalization of this argument shows that in fact any pair of distinguishable quantum states must have zero overlap~\cite{pusey2012reality}, proving that the quantum state must be ontic, modulo the assumptions $(i)-(ii)$ presented above.
Arguments that loosen the preparation independence assumption $(ii)$ can be found in Refs.~\cite{spekkens2012not,Mansfield_2016,leifer2014quantum}.

\section{A Local Friendliness no-go theorem for free choices} \label{sec:LF_free_choices}
In this section we present our main results. After defining free choices in \cref{sec:defining_free_choices}, we present our main protocol and no-go theorem against the absoluteness of free choices in \cref{sec:are_free_choices_absolute}, based on the PBR scenario. We present a generalization to arbitrary choice amplitudes in \cref{sec:generalizing_choices}, and comment on the modeling of the free choices in \cref{sec:on_modeling_choices}. Finally, in \cref{sec:use_of_correlations} we take a closer look at the empirical correlations used in this scenario, and consider arguments based on scenarios other than the PBR one in \cref{sec:other_arguments_against_free_choices}.

\subsection{Defining free choices} \label{sec:defining_free_choices}
Before we present our scenario, we first provide a definition of a `free choice' or `intervention'.\footnote{See also Ref.~\cite{wiseman2022thoughtful} for a brief discussion on interventions or free choices in the context of Local Friendliness.} 
An observer Alice making a \textit{free choice} $a$ for an experimental set-up produces a variable $a$ that is initially uncorrelated with anything else in the experiment, including both the system and the apparatus.
When asking Alice about this choice, she will reply that, as far as her computational capacity goes, she did not have knowledge of the choice's outcome before making the choice. 
In this way, we take a free choice to be whatever variable the protocol assigns the role of an observer's input required for an input setting in an experiment.
In the protocol below, Alice and Bob will have to make free choices at the start; we could, for example, think of instructing them to make free choices, and fooling them to make them believe their choices will be used in a Bell 
experiment to determine the measurement settings (instead, Wigner will perform a different operation upon them after they have made their choices).
For Wigner to model this process unitarily, we assume that the complete physical mechanism responsible for generating and recording Alice's and Bob's choices is contained within their respective laboratories.
The question of defining `free choice' can thus be argued to be related to that of defining an observer, investigated in upcoming work~\cite{walleghem2026observer}.

\subsection{Are free choices absolute?} \label{sec:are_free_choices_absolute}
\begin{figure}[h]
         \centering
\includegraphics[width=0.48\textwidth]{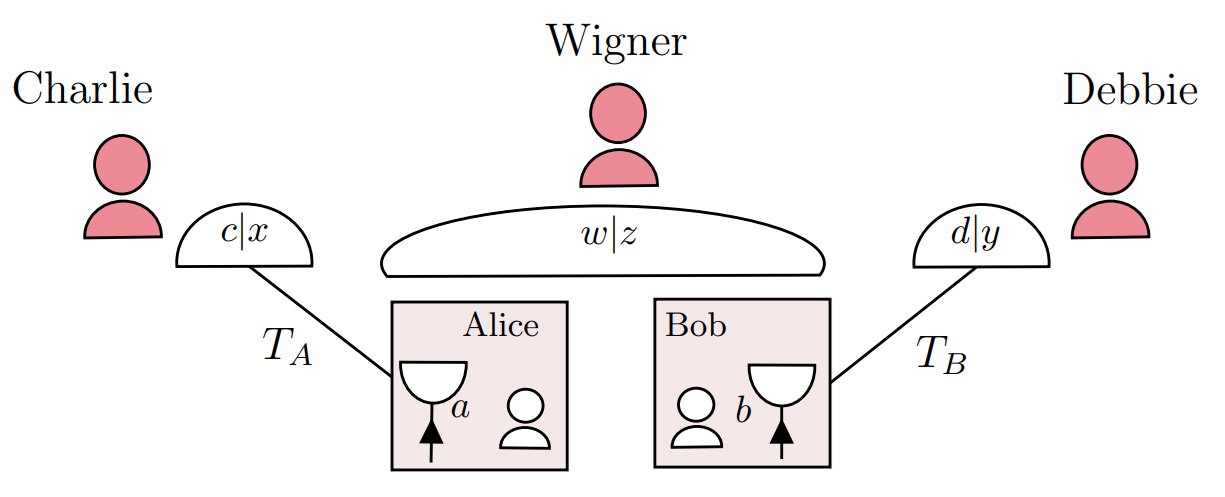}
         \caption{Protocol; 
          Alice and Bob each make a choice $a,b \in \{ 0, 1\}$. They encode their choices in the computational basis of qubits that they send to Charlie and Debbie, who choose $x,y$ whether to decode their outcomes or measure their received qubit in the $X$-basis. Wigner, a superobserver for Alice and Bob, chooses $z=0,1$. For $z=0$ he asks Alice and Bob for their choices while for $z=1$ he measures Alice's and Bob's labs jointly in an entangled PBR measurement. The only events unavailable after the protocol are the `free' choices $a,b$ of Alice and Bob.
         }
         \label{fig:PBR_LCF_scenario}
\end{figure}

The scenario that we will consider, sketched in \Cref{fig:PBR_LCF_scenario}, combines Wigner's friend with the Pusey--Barrett--Rudolph argument to derive a contradiction with the Local Friendliness assumptions, now applied to free choices.

Two friends, Alice and Bob, residing in sealed labs and modeled as quantum systems by their superobserver Wigner, each make a `free choice' $a,b \in \{ 0, 1 \}$. 
For now, we assume that Alice and Bob choose $0$ and $1$ with equal probabilities $1/2$.\footnote{In fact, more precisely, we assume for now that the amplitudes of their choices are given by $1/\sqrt{2}$.} 
They next encode their choices $a,b$ in the computational bases of qubits $T_A,T_B$ that they send to Charlie and Debbie, respectively.\footnote{The systems $T_A,T_B$ are the only systems that can escape Alice's and Bob's otherwise sealed labs.} 

Modeling the actions of Alice and Bob inside their sealed labs unitarily as viewed from the outside, we can describe the state at this point by \begin{equation}
    \label{eq:init_state_protocol_1} 
    \ket{\Phi} = \ket{\phi_0}_{T_A L_A} \otimes \ket{\phi_0}_{T_B L_B},
\end{equation} with, for $I=A,B$,\footnote{Here and later, for ease of notation, we omit the tensor product, i.e. we write $\ket{00}$ or $\ket{0}\ket{0}$ for $\ket{0}\otimes \ket{0}$.} \begin{equation}
    \label{eq:init_state_protocol_1_each_part}
        \ket{\phi_0}_{T_I L_I} = \sqrt{\frac{1}{2}} \bigg( \ket{00}_{T_I L_I} + \ket{11}_{T_I L_I} \bigg).
\end{equation}
Here $\ket{0}_{L_A}$ and $\ket{1}_{L_A}$ denote the states of Alice's lab when she chose $a=0$ and $a=1$, respectively, and similarly for Bob.

After this, Wigner chooses $z=0,1$, where for $z=0$ he reveals Alice's and Bob's choices (for example by measuring their labs in the computational basis) and for $z=1$ he performs the entangled PBR measurement on Alice's and Bob's labs $L_A,L_B$, namely measuring the composite system $L_A L_B$ in the basis $\ket{\xi_1}$, $\ket{\xi_2}$, $\ket{\xi_3}$, $\ket{\xi_4}$. 
Wigner's outcomes are denoted by $w \in \{ \xi_1, \xi_2, \xi_3, \xi_4 \}$.

Having received the qubits $T_A, T_B$ from Alice and Bob, Charlie and Debbie choose measurement settings $x$ and $y$, respectively. 
For $x=0$, Charlie decodes Alice's choice by measuring $T_A$ in the computational basis, obtaining outcome $c \in \{0,1\}$. 
For $x=1$, Charlie measures $T_A$ in the $X$-basis, obtaining outcome $c \in \{+,-\}$, and similarly for Debbie choosing $y=0,1$ to measure $T_B$ in the $Z$-- or $X$--basis.
Charlie and Debbie do so spacelike separated from each other and from Wigner. Note that Charlie and Debbie are not superobservers for Alice and Bob; they simply measure the qubit they received from Alice and Bob.

Compared to usual EWF arguments, Wigner's operations involve no `undoing', i.e. applying the inverse unitary. Instead, the records of the free choice of Alice become unavailable if neither Charlie nor Wigner reveal them, i.e. if both Charlie and Wigner measure the involved records in bases incompatible with the basis in which Alice's choice is encoded, and similarly for Bob.\footnote{The fact that no undoing is required may simplify experimental implementations, but the joint PBR measurement by Wigner remains challenging, involving projecting on states incompatible with the choice records.}

\vspace{0.5cm}

We next derive a contradiction of this scenario with the Local Friendliness assumptions. 
Absoluteness of Observed Events will be applied to the conjunction $a,b,c,d,w|x,y,z$, but the only two events unavailable (in some rounds) after the protocol are Alice's and Bob's free choices $a,b$. 
Thus, we need the assumption that the freely chosen settings $a,b$ of Alice and Bob, even when unavailable after the protocol, are absolute and single-valued.
All other measurement outcomes and choice settings are available after the protocol. 
Local Agency will be applied on various occasions in the proof to argue that certain joint variables should be uncorrelated with some of the free choices $x,y,z$ when they are outside their lightcones.

\vspace{0.5cm}

A violation of Local Friendliness is then obtained as follows. 
If $a=0,b=0$ and $x=0,y=0$, then the choices of Alice and Bob are revealed by Charlie and Debbie\footnote{To argue that, when $x=0$, Charlie's result of decoding Alice's choice from the qubit $T_A$ ís actually Alice's choice, i.e. $a = c$ when $x=0$, an additional ``tracking assumption''~\cite{schmid2023review} is introduced, and similarly for Debbie and Bob; for a discussion on why this is considered a minimal assumption, see~\cite{schmid2023review}.} and the probabilities of Wigner for $z=1$ are those of performing the PBR measurement on the state $\ket{0}\otimes\ket{0}$, as we can see from the amplitude  \begin{equation}
    \begin{split}
        \bra{0}_{T_A} \bra{0}_{T_B} \bra{\xi_i}_{L_A L_B} \ket{\phi_0}_{T_A L_A} \ket{\phi_0}_{T_B L_B} \\ = \bra{\xi_i} \ket{00}_{L_A L_B} / 2 ,
    \end{split}
\end{equation} with $\ket{\phi_0}$ given in \cref{eq:init_state_protocol_1_each_part}. Using that $\ket{\xi_1}$ is orthogonal to $\ket{0}\otimes \ket{0}$, we thus find that \begin{equation}
    \label{eq:PBR_LCF_1}  \wp(a{=}0,b{=}0,w{=}\xi_1 | x{=} 0, y{=} 0,z{=}1) = 0.
\end{equation}
If $x=0,y=1$ and $a=0,d=+$, then Charlie reveals Alice's choice by measuring the qubit $T_A$ he received from Alice in the computational basis, which turns out to be $a=0$, while Bob's choice is not revealed by Debbie who measured $T_B$ in the $X$-basis obtaining $d=+$.
Therefore, if $a=0$ and $x=0,y=1,d=+$, then the probabilities of Wigner for $z=1$ are those of performing the PBR measurement on $\ket{0}\otimes\ket{+}$, finding that 
\begin{equation}
        \label{eq:PBR_LCF_2}  \wp(a{=}0,w{=}\xi_2 | x{=} 0, y{=} 1,d{=}+,z{=}1) = 0.
\end{equation}
Similarly, if $b=0$ and $x=1,y=0,z=1,c=+$, the probabilities of Wigner are those of performing the PBR measurement on $\ket{+}\otimes\ket{0}$, finding that 
\begin{equation}
        \label{eq:PBR_LCF_3}  \wp(b{=}0,w{=}\xi_3 | x{=} 1, y{=} 0,c{=}+,z{=}1) = 0.
\end{equation}
Finally, if $x=1,y=1,z=1,c=+,d=+$, then the probabilities of Wigner are those of performing the PBR measurement on the $\ket{+}\otimes\ket{+}$ state, finding 
\begin{equation}
        \label{eq:PBR_LCF_4}  \wp(w{=}\xi_4 | x{=} 1, y{=} 1,c{=}+,d{=}+,z{=}1) = 0.
\end{equation}

Next, we will use the Local Friendliness assumptions. In this case, AOE and Local Agency imply that $p(a,b,w|x,y,z)=p(a,b,w|z)$, i.e. the joint correlations of $a,b,w$, which always exist by AOE, should be uncorrelated with $x$ and $y$ as $a,b,w$ are outside the future lightcones of $x$ and $y$. Moreover, they also imply that $p(a,b,w,c|x,y,z)=p(a,b,w,c|x,z)$ and $p(a,b,w,d|x,y,z)=p(a,b,w,d|y,z)$.

Applying AOE and Local Agency to \cref{eq:PBR_LCF_1}, we obtain that 
\begin{equation}
    \label{eq:PBR_LCF_1_1} p(a{=}0,b{=}0,w{=}\xi_1 | x{=} 1, y{=} 1,z{=}1) = 0,
\end{equation} and, using $p(c{=}+,d{=}+|x{=}1,y{=}1,z{=}1) > 0$, that 
\begin{equation}
    \label{eq:PBR_LCF_1_final}  p(a{=}0,b{=}0,w{=}\xi_1 | x{=} 1, y{=} 1, c{=} +, d{=}+,z{=}1) = 0.
\end{equation}
Similarly, applying AOE and Local Agency to \cref{eq:PBR_LCF_2}, we obtain that
\begin{equation}
        \label{eq:PBR_LCF_2_2}  p(a{=}0,w{=}\xi_2 | x{=} 1, y{=} 1,d{=}+,z{=}1) = 0,
\end{equation} and, using $p(c{=}+,d{=}+|x{=}1,y{=}1,z{=}1) > 0$, that
\begin{equation}
        \label{eq:PBR_LCF_2_final}  p(a{=}0,b{=}0,w{=}\xi_2 | x{=} 1, y{=} 1,c{=}+,d{=}+,z{=}1) = 0.
\end{equation}
The same reasoning for \cref{eq:PBR_LCF_3} leads to
\begin{equation}
        \label{eq:PBR_LCF_3_final}  p(a{=}0,b{=}0,w{=}\xi_3 | x{=} 1, y{=} 1,c{=}+,d{=}+,z{=}1) = 0.
\end{equation}

Summarizing, we have the following four conditions (from \cref{eq:PBR_LCF_1_final,eq:PBR_LCF_2_final,eq:PBR_LCF_3_final,eq:PBR_LCF_4}) on $p(a,b,w|E)$ with $E:=(x{=}1,y{=}1,c{=}+,d{=}+,z{=}1)$:
\begin{equation} \label{eq:PBR_LF_together}
    \begin{split}
         p(a=0,b=0,w=\xi_1 | E) = 0, \\
         p(a=0,b=0,w=\xi_2 |E) = 0, \\
          p(a=0,b=0,w=\xi_3 | E) = 0, \\
          p(a=0,b=0,w=\xi_4 | E) = 0.
    \end{split}
\end{equation}
We thus arrive at the following: \begin{equation} p(a{=}0,b{=}0,c{=}+,d{=}+|x{=}1,y{=}1,z{=}1)=0. \end{equation}
However, this cannot be true, as by Local Agency we have that $p(a,b,c,d|x,y,z)=p(a,b,c,d|x,y)$ and \begin{equation}
    \wp(a{=}0,b{=}0,c{=}+,d{=}+|x{=}1,y{=}1,z{=}0) > 0.
\end{equation}
We thus obtain a contradiction, finding a violation of the conjunction of Local Agency and the assumption that the events involving Alice's and Bob's free choices, being the only unavailable events at the end of the protocol, are absolute and single-valued.

\subsection{Generalizing to arbitrary choice amplitudes for Alice and Bob} \label{sec:generalizing_choices}
In the protocol outlined above, we have assumed that Alice chooses $a=0,1$ with equal probabilities, or, more correctly, with equal amplitudes $1/\sqrt{2}$, leading to the state in \cref{eq:init_state_protocol_1_each_part} after her encoding of her choice in $T_A$, and similarly for Bob. 
In general, Alice could choose $a=0,1$ with arbitrary probabilities and amplitudes, leading to \begin{equation} \label{eq:psi_choices_remark}
    \ket{\psi}_{T_A L_A} = \alpha \ket{00}_{T_A L_A}+\beta \ket{11}_{T_A L_A}.
\end{equation}
Then, to obtain a contradiction, we proceed as follows. 
We let Alice and Bob make their choices several times, while Wigner performs state tomography on Alice's and Bob's labs in calibration runs when they have made their choices. This allows to determine $\alpha$ and $\beta$. To next continue the protocol, there are two possibilities. 

The first possibility is that for $x=0$, Charlie still decodes Alice's choice by measuring $T_A$ in the computational basis, whereas for $x=1$ he now measures $T_A$ in a basis that contains the states \begin{equation}
    \ket{\Gamma} = \gamma \ket{0} + \delta \ket{1},
\end{equation} and a state $\ket{\lnot \Gamma}$ orthogonal to $\ket{\Gamma}$ with 
\begin{equation} \gamma^* \alpha = \delta^* \beta. \end{equation}
In this way, when Charlie obtains $c=\Gamma$, the post-selected state for $L_A$ is $\ket{+}$. Similarly for Debbie choosing $y$. Wigner acts as before, and a contradiction is obtained in the same way as before.\footnote{Strictly speaking, this construction works in all cases except when Alice always chooses the same setting,
i.e. $|\alpha| \notin \{0,1\}$.}

The second possibility is that Charlie and Debbie still measure in the $Z$- or $X$-basis, but that Wigner performs a PBR measurement that discriminates between $\ket{0}$ and $\alpha \ket{0} + \beta \ket{1}$. For this, we actually need several additional agents to Alice and Bob and Charlie and Debbie. Namely, we need $n$ agents making their choice and encoding their choice to send to some faraway observers who measure in the $Z$- or $X$-basis, with $n$ determined by the general joint PBR measurement as in the original PBR argument~\cite{pusey2012reality}.

\subsection{On modeling the choices of Alice and Bob} \label{sec:on_modeling_choices}
In the above argument, regarding state tomography by Wigner, we have assumed that each time Alice and Bob make a choice, which they encode in the computational bases of the qubits $T_A,T_B$, the same coefficients $\alpha$ and $\beta$ appear in \cref{eq:psi_choices_remark}. 
The motivation for this is that, assuming we initialized the state of Alice's lab in each run in exactly the same state, and take the same external conditions, the unitary evolution of the lab is the same and always leads to \cref{eq:psi_choices_remark}, similarly to repeatedly preparing a quantum random number generator. 
One could argue that this exposes some tension with the intuitive concept of a free choice.  However, this does not require the same value for Alice's choice in each run from Alice's own point of view, see also \cref{sec:discussion}.

If we instead feed in different initial conditions for Alice's lab, a slightly different environment or evolution, then we would need Charlie and Debbie to additionally keep track of this environment control in order to obtain the coefficients $\alpha$ and $\beta$ that depend on this environment. Then, in the actual protocol, when they choose their basis $\ket{\Gamma},\ket{\lnot \Gamma}$, they would need to additionally control on the environment states in order for the protocol to succeed.

\subsection{Use of correlations} \label{sec:use_of_correlations}
Our scenario can also be seen as Wigner swapping the entanglement between Alice and Charlie, and Bob and Debbie, to entanglement between Charlie and Debbie. 
The additional measurements by Charlie and Debbie lead to correlations violating Bell locality, as proved in Appendix A. 
In fact, most extended Wigner's friend arguments are based on Bell nonlocality, and, more recently, noncontextuality~\cite{szangolies2023quantum,walleghem2023extended,walleghem2024connecting,walleghem2025extendedwignersfriendnogo}.
Also the Pusey--Barrett--Rudolph theorem can be seen as capturing a form of nonclassicality or quantum advantage in some tasks~\cite{ying2026contextualadvantageconclusiveexclusion}, and is closely related to the Peres--Mermin square~\cite{karanjai2018contextualityboundsefficiencyclassical}.
One could try to devise an EWF argument based purely on the PBR set-up without these additional measurements by referring to causal or operational no fine-tuning in the setting of Wigner's friend~\cite{walleghem2023extended,ying2023relating}; we leave this for future work.

\subsection{Arguments against the absoluteness of free choices based on scenarios other than PBR} \label{sec:other_arguments_against_free_choices}
One can devise similar EWF arguments against free choices based on other noncontextuality~\cite{abramsky2011sheaf,budroni_kochen-specker_2022,spekkens2005contextuality,schmid2024noncontextualityinequalitiespreparetransformmeasurescenarios,schmid2020structure,catani2023mathematical,Chaturvedi_2020} and nonlocality scenarios. For example, such an argument resembling the minimal LF scenario~\cite{wiseman2022thoughtful} with the initial part resembling that of the original FR paradox~\cite{frauchiger2018quantum} is presented in Appendix B.

\section{Discussion} \label{sec:discussion}
We have provided a scenario that violates the Local Friendliness assumptions by combining Wigner's friend with the PBR theorem. 
The only relevant variables that are unavailable after the protocol are Alice's and Bob's free choices. Thus, under the same locality assumption as the Local Friendliness no-go theorem, this provides an argument against the absoluteness of free choices.
It is assumed that the physical mechanisms for generating and recording Alice's and Bob's free choices or interventions are included among the systems coherently controlled by their superobserver Wigner, analogously to how earlier extended Wigner's friend arguments assume all mechanisms generating the friend's measurement outcome to be included among the systems coherently controlled by their respective superobserver.

Resolutions of extended Wigner's friend paradoxes can be categorized as (i) refuting the concept of a superobserver who can model other observers unitarily, (ii) refuting absoluteness, or (iii) adhering to absoluteness at the cost of fine-tuning, see also Ref.~\cite[Chapter 10]{walleghem2026thesis} and  Refs.~\cite{walleghem2025extendedwignersfriendnogo,ying2023relating}.
Our argument thus suggests that relational resolutions to Wigner's friend paradoxes per (ii) must also incorporate a relational nature of free choices, and that resolutions per (iii) must involve fine-tuned choices in addition to fine-tuned measurement outcomes. 
It would be interesting to connect our no-go theorem to formulations of time-symmetric quantum theory that emphasize the importance of notions that connect to free choices such as priors and `agency' in physics~\cite{ormrod2026decoherencestatecausalquantum,di2021arrow,rovelli2020agencyphysics,parzygnaet2024bayes,Parzygnat2023axiomsretrodiction}.
 
\vspace{0.5cm}

One proposal for a relational resolution to extended Wigner's friend paradoxes is the Relativity of Observed Events~\cite{walleghem_refined_2024} as a replacement for the Absoluteness of Observed Events. Extending this principle to interventions, one obtains: \begin{quote}
    \textbf{Relativity of Observed Events}: Every observed event can be assigned a single value by an agent who observes that event.
\end{quote} Here `observing an event' means `learning about the value of that event through (indirect) interactions with the system holding the record of that event',  and events comprise both outcomes of performed measurements and interventions or free choices.
The Relativity of Observed Events can be phrased as an extension of Peres's dictum, \begin{quote}
    `Unperformed operations have no results, and unknown results have no values',
\end{quote} where `operations' now comprise both measurements and interventions.

\vspace{0.5cm}

Different interpretations of quantum mechanics evade extended Wigner’s friend no-go theorems by rejecting different assumptions. Objective- and gravity-induced collapse interpretations~\cite{ghirardi1986unified,bassi2003dynamical,Bassi_2013,penrose1996gravity,diosi1987universal} reject the superobserver's universal unitary description, while Bohmian approaches retain absolute outcomes but abandon the relevant locality assumption through nonlocal and contextual dynamics~\cite{lazarovici2019quantum,drezet2018wigner,sudbery2017single,sudbery2019hidden}. 
Everettian~\cite{everett1957}, QBist~\cite{fuchs2025qbismpolishingpoints,Fuchs_2014} and relational interpretations~\cite{rovelli1996relational} instead reject Absoluteness of Observed Events in different senses: outcomes are branch-relative in Everettian quantum mechanics, personal experiences in QBism~\cite{debrota2020respecting,cavalcanti2021view}, and relative to interacting physical systems in relational interpretations of quantum mechanics~\cite{cavalcanti2023consistency,di2021stable}. 
Consistent-histories~\cite{griffiths1984consistent,losada2019frauchiger} (and similarly movable-Heisenberg cut Copenhagenish interpretations~\cite{vilasini2022general,schmid2025copenhagenishinterpretationsquantummechanics}) argue that facts about measurement outcomes might be incompatible between inconsistent histories (or cuts), which can be interpreted as rejecting the cross-history assignment required by Absoluteness.

The present result is compatible with these responses,
but sharpens what they require: when the physical process producing an intervention is itself coherently controllable by a superobserver (as collapse and objective-Heisenberg-cut interpretations would expectedly deny), the intervention value itself requires reinvestigation.
Bohmian and perhaps superdeterministic interpretations~\cite{Hossenfelder_2020} would presumably retain absolute interventions while abandoning Local Agency;
Everettian, QBist, and relational approaches might instead deny absolute intervention settings. Movable-Heisenberg-cut interpretations could likewise place the interventions in our thought experiment below the cut and deny their values absolute status.
To our knowledge, existing interpretations treat interventions primarily as external inputs. 
It may therefore be particularly interesting to rethink interpretations while assigning free choice a more fundamental role.
Causal decoherence theory~\cite{ormrod2026decoherencestatecausalquantum} is one recent dynamics-first interpretation rejecting Absoluteness that emphasizes the dual roles of outcomes and quantum states, with events arising from decoherence and dual decoherence, which are relative to a set of systems of interest.

An important question for interpretations rejecting Absoluteness is \textit{why} some facts can be combined consistently, while others cannot, ideally with an explanation grounded in a concrete physical mechanism.
The aforementioned Relativity of Observed Events may contribute toward such account by grounding relative facts in (in)direct interactions through which an agent learns an event's value.

\vspace{0.5cm}

Finally, in order to devise the argument, the free choices made by Alice and Bob were modeled in a particular way that internalizes the free choice, rather than treating them as exogenous. 
They could still seem exogenous to Alice and Bob, 
but the superobserver Wigner considers them internalized.
We have tried not to subscribe to any interpretation of free choices a priori, i.e. whether free choice is an emergent concept or not, and our argument is not an argument against the notion of free choice, only against the absoluteness of it, modulo quantum universality and a locality assumption.
It might feel unnatural to model free choices in the way we did, or perhaps in any way other than exogenously from the perspective of the entity who made that choice.
Should one object to such modeling, at least, our argument invites a careful operational rethinking of free choices and their role in physics.

\vspace{0.4cm}

\section*{Acknowledgements} \vspace{-0.1cm}
I thank Lorenzo Catani for interesting discussions on the notion of free choice and for feedback on this manuscript. I thank Rui Soares Barbosa and Matt Pusey for feedback on an earlier version of this manuscript. I thank Eric Cavalcanti and Stefan Weigert for many interesting discussions on Wigner's friend. 
I acknowledge support from the United Kingdom Engineering and Physical Sciences Research Council (EPSRC) through the DTP Studentship EP/W524657/1, and thank INL and its QLOC group in Braga, Portugal for the kind hospitality.
I thank the anonymous reviewers and editor for their constructive comments.

\bibliography{refs}

\section*{Appendix A: Entanglement swapping correlations violating Bell locality} \label{sec:app_A}

The level of the contradiction in the presented Wigner's friend argument based on the PBR scenario is at the possibilistic level; the nonlocality is also present at the possibilistic level, as we show now.
We can see the obtained measurement outcomes in the Wigner's friend PBR argument as a tripartite Bell scenario, involving Charlie, Debbie and Wigner, choosing settings $x,y,z$ and obtaining outcomes $c,d,w$. The initial state is given as in \cref{eq:init_state_protocol_1}. 
Recall that Charlie chooses $x=0,1$ to measure $T_A$ in the $Z$- or $X$-basis, and similarly for Debbie measuring $T_B$. 
Wigner makes a choice $z=0,1$. For $z=1$ he performs the joint PBR measurement on $L_A  L_B$ with outcome $w \in  \{ \xi_i \}_{i=1,\ldots,4}$. For $z=0$ he reveals Alice's and Bob's choices, meaning he measures $L_A L_B$ in the computational basis. We label the corresponding outcome for $\ket{ij}_{L_A L_B}$ by $w= (ij) | z= 0$ for $i,j \in \{0,1\}$. 

\vspace{0.5cm}

The empirical probabilities that are used in the scenario are \begin{equation} \label{eq:proof_nonlocality1}
    \begin{split}
        &\wp(c{=}0,d{=}0,w{=}\xi_1|x{=}0,y{=}0,z{=}1)=0, \\
        &\wp(c{=}0,d{=}+,w{=}\xi_2|x{=}0,y{=}1,z{=}1)=0, \\
        &\wp(c{=}+,d{=}0,w{=}\xi_3|x{=}1,y{=}0,z{=}1)=0, \\
        &\wp(c{=}+,d{=}+,w{=}\xi_4|x{=}1,y{=}1,z{=}1)=0, \\
    \end{split}
\end{equation}
and \begin{equation} \label{eq:proof_nonlocality2}
    \wp(c{=}+,d{=}+,w{=} (00)|x{=}1,y{=}1,z{=}0) > 0.
\end{equation} 
Moreover, as $c|x=0$ is actually Alice's choice $a$ as revealed by Wigner when $z=0$, and similarly for Debbie and Bob, we have that \begin{equation} \label{eq:w_delta_ab}
    \wp(c,d,w|x{=}0,y{=}0,z{=}0) = \delta_{w,(c,d)} \wp(w|z{=}0).
\end{equation}
Now it is easy to see that these correlations are logically contextual (or, equivalently in this case, logically nonlocal), i.e. contextual on a possibilistic level, meaning contextuality can be obtained purely from knowing the support of the probability distributions~\cite{abramsky2011sheaf}.
To simplify stating the argument, we label the outcome $c$ when performing measurement $x=0,1$ by $c_0,c_1$, i.e., $c_x= (c|x)$ and similarly $d_y=(d|y)$ and $w_z = (w|z)$.
Then, we find from \cref{eq:proof_nonlocality2} that the assignment $(c_1{=}+,d_1{=}+,w_0{=}(00))$ is possible in the context $C=(x{=}1,y{=}1,z{=}0)$. Moreover, by \cref{eq:w_delta_ab}, $w_0=(00)$ implies that $c_0=0=d_0$. However, this assignment cannot be extended to a global assignment; one cannot assign a value to $w_1$ consistent with \cref{eq:proof_nonlocality1}, meaning the empirical correlations are logically contextual. 
We verify this as follows. 
As $c_0{=}0{=}d_0$, by the first equation in \cref{eq:proof_nonlocality1}, we cannot have $w_1 = \xi_1$. As $c_0{=}0,d_1{=}+$, we cannot have $w_1=\xi_2$ by the second equation in \cref{eq:proof_nonlocality1}. Finally, as $c_1{=}+,d_0{=}0$, by the third equation in \cref{eq:proof_nonlocality1} we cannot have $w_1=\xi_3$, and by the fourth equation and $c_1{=}+{=}d_1$ we cannot have $w_1=\xi_4$.

\section*{Appendix B: an argument against free choices in the minimal LF scenario}
In this Appendix we present an EWF argument based on a scenario other than the PBR scenario. The argument resembles the minimal LF scenario~\cite{wiseman2022thoughtful} with the initial part resembling that of the original FR paradox~\cite{frauchiger2018quantum}. 

Alice makes a free choice between $a=0,1$ with amplitudes $1/\sqrt{3}, \sqrt{2/3}$ and prepares a system $S$ in the state $\ket{0}$ or $\ket{+}$ for choices $a=0$ and $a=1$, respectively, giving the state $1/\sqrt{3}(\ket{00}+\ket{10}+\ket{11})_{A S}$. 
Alice sends the qubit $S$ to Bob, who chooses $y=0,1$ to measure it in the $Z$- or $X$-basis. Wigner, a superobserver for Alice (only), chooses $x=0,1$ whether to reveal Alice's choice or measure her in the $X$-basis. 
The empirical correlations obtained violate Bell locality and thus do not admit a global probability distribution which should exist by AOE and Local Agency, with now the only variable that is not always available after the protocol being Alice's choice $a$.
To generalize to arbitrary choice amplitudes for Alice, the measurements by Wigner and Bob should be altered.

\end{document}

%% file: preamble.tex
\usepackage{graphicx}
\usepackage{tikz}
\usetikzlibrary{shadows.blur}

\usepackage{physics}
\usepackage{braket}
\usepackage{float}  
\usepackage[english]{babel}
\usepackage[utf8]{inputenc}

\usepackage[colorlinks=true,linktocpage=true]{hyperref}
\hypersetup{
    allcolors  = {blue},
}
\usepackage{booktabs} 

\usepackage{amsmath}
\usepackage{amssymb} 
\usepackage{bbold}
\usepackage{amsthm}

\usepackage{xfrac}
\usepackage{mathtools}

\usepackage{standalone}

\usepackage{thmtools}
\usepackage{thm-restate}

\usepackage{listings}
\usepackage{enumitem}   

\usepackage[page,toc,titletoc,title]{appendix}

\usepackage{footmisc}
\usepackage[justification=justified,singlelinecheck=false]{caption}

\usepackage{microtype} 
\usepackage[capitalize]{cleveref} 
\crefname{section}{Sec.}{Secs.}

\usepackage{nomencl}
\makenomenclature

\usepackage{soul}
\newtheorem{theorem}{Theorem}

\newtheorem{assump}{Assumption} 

\usepackage{csquotes}